\documentclass[twopage,11pt] {article}  
\usepackage{latexsym}
\input amssym.tex
\usepackage{%
pstricks,%
pst-plot,%
epsf%
}                               
\usepackage{cite}
\usepackage{graphicx}
\usepackage{array,dcolumn}      
\usepackage{axodraw}
\newcommand{\hs}{\hspace*{0.5cm}}
\newcommand{\vs}{\vspace*{0.5cm}}
\newcommand{\be}{\begin{equation}}
\newcommand{\ee}{\end{equation}}
\newcommand{\bea}{\begin{eqnarray}}
\newcommand{\eea}{\end{eqnarray}}
\newcommand{\bary}{\begin{array}}
\newcommand{\eary}{\end{array}}
\newcommand{\bit}{\begin{itemize}}
\newcommand{\eit}{\end{itemize}}
\newcommand{\ben}{\begin{enumerate}}
\newcommand{\een}{\end{enumerate}}
\newcommand{\crn}{\nonumber \\}
\newcommand{\nn}{\nonumber}

\newcommand{\al}{\alpha}
\newcommand{\la}{\lambda}
\newcommand{\bet}{\beta}
\newcommand{\ga}{\gamma}
\newcommand{\va}{\varphi}

\newcommand{\om}{\omega}

\newcommand{\fr}{\frac}

\newcommand{\bc}{\begin{center}}
\newcommand{\ec}{\end{center}}
\newcommand{\Ga}{\Gamma}

\newcommand{\ep}{\epsilon}

\newcommand{\si}{\sigma}

\newcommand{\Om}{\Omega}
\begin{document}           
\title{Candidates for
dark matter   in the
 $\mbox{SU}(3)_C\otimes \mbox{SU}(3)_L \otimes \mbox{U}(1)_N$
 models }  
\author{
 H. N. Long$^a$, N. Q.  Lan$^b$,
 D. V.  Soa$^b$ and  L. N. Thuc$^b$
\vspace{0.5cm}
\\
 $^a$ {\it  Institute of  Physics, VAST,
P. O. Box 429, Bo Ho, Hanoi, Vietnam}\\
$^b${\it   Department of Physics, Hanoi University of Education,
Hanoi, Vietnam }}
\maketitle                 
\pagestyle{myheadings} \thispagestyle{plain} \markboth{H. N. Long
et al}{Candidates for Dark Matter} \setcounter{page}{1} It has
recently been pointed out that the 511 keV emission line detected
by INTEGRAL/SPI from the bulge of our galaxy could be explained by
annihilations of light dark matter particles into $e^+\ e^-$. We
present the possibility that dark matter could be made of scalar
candidates, namely , of the Higgs bosons in the models based  on
$\mbox{SU}(3)_C\otimes \mbox{SU}(3)_L \otimes \mbox{U}(1)_N$
(3-3-1) gauge group. These particles are singlet of the
$\mbox{SU}(2)_L \otimes \mbox{U}(1)_Y$ group, so they do not
interact with the ordinary
 particles, exept the Higgs boson in the standard model.
The Spergel-Steinhardt condition for self-interacting dark matter
gives a bound on the mass of the candidates to be a few MeVs.
Besides the scalar candidates, which exist in both non-SUSY and
SUSY 3-3-1 models with right-handed neutrinos,
 the spin $\fr 1 2$ candidate exists in a variant 3-3-1 version
with exotic neutral lepton.
 In contrast to the singlet models, where an extra symmetry must
be imposed to account the stability of the dark matter, here the
decay of the candidates is automatically forbidden in all orders
of perturbative expansion. This is because of the following
feature:
 these scalars are singlets, i.e., in
bottom of the Higgs  triplet. Therefore, the standard model
fermions and the standard gauge bosons cannot couple with them.
\section{Introduction}
 Observation of the cosmic microwave background, the primordial
abundances of light elements and large scale structure have
revealed that a great deal of the mass of our universe consists of
dark matter (DM). It is an amazing fact that even as our
understanding of cosmology progresses by leaps and bounds, we
remain almost completely ignorant about the nature of most of the
matter in the universe~\cite{mt00}. It has recently been pointed
out that the 511 keV emission line detected by INTEGRAL/SPI from
the bulge of our galaxy could be explained by annihilations of
light dark matter particles into $e^+\ e^-$.  The nature of DM is
still a challenging question in cosmology. Cosmological models
with a mixture of roughly 35\% collisionless cold DM such as
axions, WIMPs, or any other candidate interacting through the weak
and gravitational forces only, and 65\% vacuum energy or
quintessence match observation of the cosmic microwave background
and large scale structure on extra-galactic scales with remarkable
accuracy~\cite{bah99,wan00}. It is known that only a fraction of
the dark matter can be made of ordinary baryons and its enormous
amount has unknown, nonbaryonic origin~\cite{pos01}.
 Until a few years ago, the more satisfactory cosmological
scenarios were those ones composed of ordinary matter, cold DM and
a contribution associated with the cosmological constant. To be
consistent with inflationary cosmology, the spectrum of density
fluctuations would be nearly scale-invariant and adiabatic.
However, in recent years it has been pointed out that the
conventional models of collisionless cold DM lead to problems with
regard to galactic structures. They were only able to fit the
observations on large scales ($\gg 1$ Mpc). Also, $N$-body
simulations in these models result in a central singularity of the
galactic halos~\cite{ghi} with a large number of
sub-halos~\cite{moore}, which are in conflict with astronomical
observations. A number of other inconsistencies are discussed in
Refs.~\cite{dav01,bul00}. Thus, the cold DM model is not able to
explain observations on scales smaller than a few Mpc. It has
recently been shown that an elegant way to avoid these problems is
to assume the so called {\it self-interacting dark
matter}~\cite{ss}. One should notice that, in spite of all,
self-interacting models lead to spherical halo centers in
clusters, which is not in agreement with ellipsoidal centers
indicated by strong gravitational lens observations~\cite{yoh00}
and by Chandra observations~\cite{buote}.
 However, self-interacting dark matter models are self-motivated
as alternative models.  It is a well-accepted fact that the
plausible candidates for DM are elementary particles. The key
property of these particles is that,  they must  have a large
scattering  cross-section and negligible annihilation or
dissipation. The Spergel-Steinhard  model  has motivated many
follow-up studies~\cite{pos01,ost,mcd}. Several authors have
proposed models in which a specific scalar singlet that satisfies
the self-interacting dark matter properties is introduced in the
standard model (SM) in an {\it ad hoc} way \cite{pos01,mcd}.
 The SM offers no options for DM. The first gauge model for SIDM
is the extended SM based on the  $\mbox{SU}(3)_C\otimes
\mbox{SU}(3)_L \otimes \mbox{U}(1)_N$  (3-3-1) gauge
group~\cite{ppf,flt}. The 3-3-1 model  founded by Fregolente and
Tonasse~\cite{ft} to be SIDM differs from another one founded by
Long and Lan~\cite{longlan}. Models with $SU(3)_C\otimes
SU(3)_L\otimes U(1)_N$ gauge symmetry (called 3-3-1 models for
short) are interesting possibilities for the physics at the TeV
scale~\cite{ppf,flt}. At low energies they coincide with the
standard model and some of them give at least partial explanation
to some fundamental questions that are accommodated but not
explained by the standard model. For instance, in order to cancel
the triangle anomalies, together with asymptotic freedom in QCD,
the model predicts that the number of generations must be three
and only three; {\it (ii)} the model of Ref.~\cite{ppf} predicts
that $(g'/g)^2=\sin^2\theta_W/(1-4\sin^2\theta_W)$, thus there is
a Landau pole at the energy scale $\mu$ at which
$\sin^2\theta_W(\mu)=1/4$.  According to recent calculations
$\mu\sim4$ TeV~\cite{phf,pl331} ; {\it (iii)} the quantization of
the electric charge~\cite{pr} and the vectorial cha\-rac\-ter of
the electromagnetic interactions~\cite{cp} do not depend on the
nature of the neutrinos i.e., if they are Dirac or Majorana
particles; {\it (iv)} as a consequence of item ii) above, the
model possesses perturbative ${\cal N}=1$ supersymmetry naturally
at the $\mu$ scale~\cite{331susy,marcos}; {\it (v)} the
Peccei-Quinn~\cite{pq} symmetry occurs naturally in these
models~\cite{pal}; {\it (vi)} since one generation of quarks is
treated differently from the others this may be lead to a natural
explanation for  the large mass of the top quarks~\cite{longvan}.
Moreover, if right-handed neutrinos are considered transforming
non-trivially~\cite{flt}, 3-3-1 models~\cite{ppf} can be embedded
in a model with 3-4-1 gauge symmetry in which leptons transform as
$(\nu_l,l,\nu^c_l,\,l^c)_L\sim({\bf1},{\bf4},0)$ under each gauge
factors~\cite{f,pp95}. The $SU(3)_L$ symmetry is possibly the
largest symmetry  involving the known leptons (and $SU(4)_L$ if
right-handed neutrinos do really exist). This make 3-3-1 or 3-4-1
models interesting by their own, and it has been the source of
interest recently~\cite{rc} because it requires that the number of
fermion families be a multiple of the quark color in order to
cancel anomalies, which suggest a path to the solution of the
flavor problem. The Peccei-Quinn~\cite{pq} symmetry occurs
naturally in these models~\cite{pal}.  Another important feature
of these models is that the $\mbox{SU}(2)_L $ group is totally
embedded in $ \mbox{SU}(3)_L $.
 A subject that has not been given much attention by particle
physicists in the past, could prove to be a remarkable powerful
and precise probe of the properties of dark matter. This paper is
organized as follows: In Sec. 2 we briefly introduce necessary
elements of the non-SUSY 3-3-1 models, and the candidates for dark
matter.
 Sec. 3 is devoted to supersymmetric 3-3-1 model with right-handed
neutrinos. We summarize our result and make conclusions in the
last section.
\hs The aim of this reviewr is to show that
 the   3-3-1 models
 contains  candidates for  dark matter.  The outline of this paper
is as follows. In Sec. 3, the 3-3-1 models are briefly recalled.
Properties of DM are introduced. The candidates for SIDM in four
3-3-1 models are explored. The last section is devoted for
conclusions.
\section{The non-SUSY  3-3-1 models}
 To frame the context, it is appropriate to recall briefly some
relevant features of the 3 - 3 - 1 models\cite{ppf,flt}. We first
introduce the minimal version proposed by Pisano, Pleitez and
Frampton~\cite{ppf}
\subsection{The  3-3-1 model with exotic lepton}
 The model treats the leptons as  $\mbox{SU(3)}_L$
triplets~\cite{pt93} \be f^{a}_L = \left( \begin{array}{c}
             \nu^a_L\\   e^a_L\\   P^a
               \end{array}  \right) \sim (1, 3, 0),\  e^a_R\sim
(1, 1, -1),\  P^a_R \sim (1, 1, +1), \label{l} \ee where $ a = 1,
2, 3$ is the generation index.
 Each charged left-handed fermion field has its right-handed countepart
transforming as a singlet of the SU(3)$_L$ group.
  Two of the three quark generations transform as antitriplets and
the third generation is treated differently - it belongs to a
triplet: \be Q_{iL} = \left( \begin{array}{c}
                d_{iL}\\ -u_{iL}\\ D_{iL}\\
                \end{array}  \right) \sim (3, \bar{3}, -\frac{1}{3}),
\label{q} \ee
\[ u_{iR}\sim (3, 1, 2/3), d_{iR}\sim (3, 1, -1/3),
D_{iR}\sim (3, 1, -4/3),\ i=1,2,\] \be
 Q_{3L} = \left( \begin{array}{c}
                 u_{3L}\\ d_{3L}\\ T_{L}
                \end{array}  \right) \sim (3, 3, 2/3),
\ee
\[ u_{3R}\sim (3, 1, 2/3), d_{3R}\sim (3, 1, -1/3), T_{R}
\sim (3, 1, 5/3).\] Of the nine gauge bosons  $W^a (a = 1, 2, ...,
8)$ and $B$ of $SU(3)_L$ and $U(1)_N$, four are light: the photon
($A$), $Z$ and $W^\pm$. The remaining five correspond to new heavy
gauge bosons $Z_2,\ Y^\pm$ and the doubly charged bileptons
$X^{\pm \pm}$.
  Symmetry breaking and fermion mass generation can be
achieved by three scalar  $\mbox{SU(3)}_L$ triplets \bea
&\mbox{SU}(3)_{C}&\hspace*{-0.2cm}\otimes \
\mbox{SU}(3)_{L}\otimes \mbox{U}(1)_{N}\crn &\downarrow
&\hspace*{-0.8cm}\langle  \chi \rangle   \crn
&\mbox{SU}(3)_{C}&\hspace*{-0.2cm}\otimes \
\mbox{SU}(2)_{L}\otimes \mbox{U}(1)_{Y}\crn &\downarrow
&\hspace*{-0.8cm}\langle \eta \rangle, \langle
\rho \rangle\nonumber   \\
&\mbox{SU}(3)_{C}&\hspace*{-0.2cm}\otimes \ \mbox{U}(1)_{Q},
\nonumber \label{ssb2} \eea where the minimally required scalar
multiplets are summarized as \bea \chi & = &\left(
\begin{array}{c}
                \chi^-\\ \chi^{--}\\ \chi^0\\
                \end{array}  \right) \sim (1, 3, -1),\crn
\label{mh1} \eta & =& \left( \begin{array}{c}
                \eta^0\\ \eta_1^-\\ \eta_2^+\\
                \end{array}  \right) \sim (1, 3, 0),\crn
\rho & =& \left( \begin{array}{c}
                \rho^+\\ \rho^0\\ \rho^{++}\\
                \end{array}  \right) \sim (1, 3, +1),\crn
\eea The vacuum expectation value (VEV) $\langle \chi^T \rangle =
( 0, 0, w/ \sqrt{2})$ yields masses for the exotic quarks, the
heavy neutral gauge boson ($Z_2$) and two charged gauge bosons
($X^{++}, Y^+$). The masses of the standard gauge bosons and the
ordinary fermions are related to the VEVs of the other scalar
fields, $\langle \eta^0 \rangle = v/ \sqrt{2}, \langle \rho^o
\rangle = u/ \sqrt{2}$.
 In order to be consistent with the low energy phenomenology
we have to assume that $u \gg\ v,\ \om$.
  By matching the gauge coupling constants we get a relation
between $g$ and $g_N$ -- the couplings associated with
$\mbox{SU(3)}_L$ and $\mbox{U(1)}_N$, respectively: \be
\frac{g_N^2}{g^2} = \frac{6 \ s^2_W(M_{Z_2})}{1 - 4
s^2_W(M_{Z_2})}, \label{coupm} \ee where $e = g\  s_W$ is the same
as in the SM. The most {\it economical}, gauge invariant and
renormalizable Higgs potential is \bea V\left(\eta, \rho,
\chi\right) & = & \mu_1^2\eta^\dagger\eta +
\mu_2^2\rho^\dagger\rho + \mu_3^2\chi^\dagger\chi +
a_1\left(\eta^\dagger\eta\right)^2 +
a_2\left(\rho^\dagger\rho\right)^2 +
a_3\left(\chi^\dagger\chi\right)^2\cr &&+
\left(\eta^\dagger\eta\right)\left(a_4\rho^\dagger\rho +
a_5\chi^\dagger\chi\right) +
a_6\left(\rho^\dagger\rho\right)\left(\chi^\dagger\chi\right) +
a_7\left(\rho^\dagger\eta\right)\left(\eta^\dagger\rho\right)\cr
&& + a_8\left(\chi^\dagger\eta\right)\left(\eta^\dagger\chi\right)
+ a_9\left(\rho^\dagger\chi\right)\left(\chi^\dagger\rho\right) +
\frac{1}{2}\left(f\ep^{ijk}\eta_i\rho_j\chi_k + {\mbox{H.
c.}}\right). \label{pot}\eea
 Here the $\mu$'s and $f$ are coupling constants with dimension
of mass with $a_3 < 0$ and $f < 0$ from the positivity of the
scalar masses \cite{TAKL}.\par Symmetry breaking is initiated when
the scalar neutral fields are shifted as $\va= v_\va + \xi_\va +
i\zeta_\va$, with $\va$ $=$ $\eta^0$, $\rho^0$, $\chi^0$. The
details of the physical spectrum of the neutral scalar sector are
crucial for our results. It is given in Refs. \cite{TAKL} and we
summarize them here. Firstly we notice that real part of the
shifted fields leads to the three massive physical scalar fields
$H_1^0$, $H_2^0$, $H_3^0$ defined by \be \pmatrix{\xi_\eta \cr
\xi_\rho} \approx \frac{1}{v_W}\pmatrix{v & u \cr u &
-v\cr}\pmatrix{H^0_1 \cr H^0_2 \cr}, \qquad \xi_\chi \approx
H_3^0, \label{3H0} \ee where we are using $w \gg v, u$. The scalar
$H^0_1$ is the one that we can identify with the standard  model
Higgs, since its squared mass, \be m_1^2 \approx 4\frac{a_2u^4 -
a_1v^4}{v^2 - u^2}, \label{h1} \ee
 carries no any feature from the 3-3-1
breakdown to the standard model. On the other hand $H^0_3$, with
squared mass \be m_3^2 \approx -4a_3w^2, \label{h3} \ee is a
typical 3-3-1 scalar. So, there is no any massless Goldstone boson
rising from the real part of the neutral sector. On the other
hand, from the imaginary part we have two Goldstone and one
massive physical state $h^0$ with eigenstate \be \zeta_\chi
\approx h^0 \label{zetac} \ee and squared mass \be m_h^2 =
-f\frac{v_W^2w^2 + v^2u^2}{vuw}. \label{mh} \ee It is important to
notice that $\zeta_\eta$ and $\zeta_\rho$ are pure massless
Goldstone states. The approximation in Eqs. (\ref{3H0}),
(\ref{h1}), (\ref{h3}) and (\ref{zetac}), is valid for $w \gg v,
u$. This condition leads to relations among the parameters of the
scalar potential (\ref{pot}). One of them, which enters in the
$H_1^0h^0h^0$ interaction, is \be a_5v^2 + 2a_6u^2 \approx
-\frac{vu}{2} \label{ll} \ee
 We must consider also the matter coupling through the scalar
fields. In the model of Ref. \cite{pt93}, the full Yukawa
Lagrangians that must be considered are
 \bea {\cal L}_\ell & = &
-\sum_{ab}\left(\frac{1}{2}\epsilon^{ijk}G^{\left(\nu\right)}_{ab}
\overline{{\psi_{aiL}}^C}\psi_{bjL}\eta_k +
G^{\left(\ell\right)}_{ab}\overline{\psi}_{aL}\ell^-_{bR}\rho -
G^{\left(P\right)}_{ab}\overline{\psi }_{aL}P^+_{bR}\chi\right)
\label{yukl} + {\mbox{H. c.}}, \\
{\cal L}_Q & = &
\overline{Q}_{1L}\sum_b\left(G^{\left(U\right)}_{1b}U_{bR}\eta +
G^{\left(D\right)}_{1b}D_{bR}\rho +
G^{\left(J\right)}J_{1R}\chi\right) + \cr &&
\sum_\al\overline{Q}_{\al L}\left(F^{\left(U\right)}_{\al
b}U_{bR}\rho^* + F^{\left(D\right)}_{\al b}D_{bR}\eta^* +
\sum_{\bet}F^{\left(J\right)}_{\al\bet}J_{\bet R}\chi^*\right) +
\mbox{H. c.}, \label{yukq2}\eea where $G^{\left(\nu\right)}_{ab}$,
$G^{\left(\ell\right)}_{ab}, G^{\left(P\right)}_{ba}$,
$G^{\left(U\right)}_{1b}$, $F^{\left(U\right)}_{\al b}$,
$G^{\left(D\right)}_{1b}$, $F^{\left(D\right)}_{\al b}$,
$G^{\left(J\right)}$ and $F^{\left(J\right)}_{\al\bet}$ are the
Yukawa coupling constants. $\eta^*$, $\rho^*$ and $\chi^*$ denote
the $\eta$, $\rho$ and $\chi$ antiparticle fields, respectively.
 The main properties that a good dark matter candidate must
satisfy are stability and neutrality. Therefore, we go to the
scalar sector of the model, more specifically to the neutral
scalars, and we examine whether any of them can be stable and in
addition whether they can satisfy the self-interacting dark matter
criterions \cite{ss}. In addition, one should notice that such
dark matter particle must not overpopulate the Universe. On the
other hand, since our dark matter particle is not imposed
arbitrarily to solve this specific problem, we must check that the
necessary values of the parameters do not spoil the other bounds
of the model.\par We can check through a direct calculation by
employing the Lagrangians (\ref{pot}), (\ref{yukl}) and
(\ref{yukq2}) and by using the eigenstates (\ref{3H0}) and
(\ref{zetac}) that the Higgs scalar $h^0$ and $H^0_3$ can, in
principle, satisfy the criterions above. Remarkably they do not
interact directly with any standard model field except for the
standard Higgs $H^0_1$. However, $h^0$ must be favored, since we
have checked that it is easier to obtain a large scattering cross
section for it, relative to $H^0_3$, by a convenient choice of the
parameters.
 In contrast to the singlet models of the
 Refs.\cite{pos01,mcd}, where an extra symmetry must be imposed to
account the stability of the dark matter, here the decay of the
$h^0$ scalar is automatically forbidden in all orders of
perturbative expansion. This is because of the following features:
i) this scalar comes from the triplet $\chi$, the one that induces
the spontaneous symmetry breaking of the 3-3-1 model to the
standard model. Therefore, the standard model fermions and the
standard gauge bosons cannot couple with $h^0$, ii) the $h^0$
scalar comes from the imaginary part of the Higgs triplet $\chi$.
As we mentioned above, the imaginary parts of $\eta$ and $\rho$
are pure massless Goldstone bosons. Therefore, there is not
physical scalar fields which can mix with $h^0$. So, the only
interactions of $h^0$ come from the scalar potential and they are
$H_3^0h^0h^0$ and $H_1^0h^0h^0$. This latter has strength
$2i\left(a_5v^2 + a_6u^2\right)/v_W \equiv 2i\Theta$. We can check
 also that $h^0$ does not interact with other exotic particle.
\par Hence, if $v \sim u \sim (100 - 200)$ GeV and $-1 \leq a_5
\sim a_6 \leq 1$, the $h^0$ can interact only weakly with ordinary
matter through the Higgs boson of the standard model $H_1^0$. The
relevant quartic interaction for scattering is $h^0h^0h^0h^0$,
whose strength is $-ia_3$. Other quartic interactions evolving
$h^0$ and other neutral scalars are proportional to $1/w$ and so
we neglect them. The cross section of the process $h^0h^0 \to
h^0h^0$ {\it via} the quartic interaction is $\si = a_3^2/64\pi
m_h^2$. The contribution of the trilinear interactions {\it via}
$H_1^0$ and $H_3^0$ exchange are negligible. There is no other
contribution to the process involving the exchange of vector or
scalar bosons. A self-interacting dark matter candidate must have
mean free path $\Lambda = 1/n\si$ in the range 1 kpc $< \Lambda <$
1 Mpc, where $n = \rho/m_h$ is the number density of the $h^0$
scalar and $\rho$ is its density at the solar radius \cite{moore}.
Therefore, with $a_3!
 = -1$, $-0.208 \times 10^{-7} {\mbox{ GeV}}
 \leq f \leq -0.112 \times 10^{-6} {\mbox{ GeV}}$,
 $w = 1000$ GeV, $u = 195$ GeV and $\rho = 0.4$ GeV/cm$^3$,
 we obtain the required Spergel-Steinhardt bound, {\it i. e.},
 $2 \times 10^3 {\mbox{ GeV}}^{-3} \leq \si/m_h \leq 3
 \times 10^4 {\mbox{ GeV}}^{-3}$ \cite{ss}.\par
With this set of parameter values, we see from Eq. (\ref{mh}) that
5.5 MeV $\leq m_h \leq$ 29 MeV. This means that our dark matter
particle is non-relativistic in the decoupling era (decoupling
temperature $\sim$ 1 eV) and, for a standard model Higgs boson
mass $\sim 100$ GeV \cite{pdg}, it is produced by a thermal
equilibrium density of the standard Higgs scalar to $h^0h^0$ pairs
\cite{mcd}. The density of the $h^0$ scalar from the $H^0_1$ decay
can be obtained following the standard procedure, {\it i. e.}, we
must solve the Boltzmann equation \be \frac{dn_h}{dt} + 3Hn_h =
\langle\Ga_H\rangle n^{({\rm eq})}_H, \label{bolt}\ee where $n_h$
is the number density of the $h^0$ scalar at the time $t$, $H$ is
the Hubble expansion rate, \be \Ga_H = \frac{\Theta^2}{4\pi E} \ee
is the decay rate for the $H_1^0$ with energy $E$ and \be n^{({\rm
eq})}_H = \frac{1}{2\pi^2}\int^\infty_{m_1}{\frac{E\sqrt{E^2 -
m_1^2}}{{\rm e}^{E/T} - 1}} \ee is the thermal equilibrium density
of the standard $H^0_1$ at temperature $T$ \cite{kt}. We are using
the condition that the temperature is less than the electroweak
phase transition $T_{\rm EW} \geq 1.5m_1$ \cite{mcd}. The thermal
average of the decay rate is given by \be \langle\Ga_H\rangle =
\frac{\al \left(\Theta T\right)^2}{8\pi^3n_H^{({\rm eq})}}{\rm
e}^{m_1/T}, \ee where $\al$ is an integration parameter that can
be taken to be 1.87 \cite{mcd}. We define $\bet \equiv n_h/T^3$
and in the radiation-dominated  era we write the evolution
equation (\ref{bolt}) as \be \frac{d\bet}{dT} =
-\frac{\langle\Ga_H\rangle\bet^{({\rm eq})}}{KT^3} =
-\frac{\al}{8\pi^3K{\rm
e}^{m_1/T}}\left(\frac{\Theta}{T^2}\right)^2, \ee where $K^2 =
4\pi^3g\left(T\right)/45m_{\rm Pl}^2$, $\bet^{({\rm eq})} =
n^{({\rm eq})}_h/T^3$ is the $\bet$ parameter in the thermal
equilibrium, $m_{\rm Pl} = 1.2 \times 10^{19}$ GeV is the Planck
mass and $g\left(T\right) = g_B + 7g_F/8 = 136.25$ for the model
of the Ref. \cite{pt93}. $g_B$ and $g_F$ are the relativistic
bosonic and fermionic degrees of freedom, respectively. Here we
are taking $T = m_1$ since this regime gives the larger
contribution to $\bet$ \cite{mcd}. Hence, \be \bet = \frac{\al
\Theta^2}{4 \pi^3Km_1^3}. \label{beta}\ee Now, the cosmic density
of the $h^0$ scalar is \be \Om_h =
2g\left(T_\Ga\right)T^3_\ga\frac{m_h\bet}{\rho_cg\left(T\right)},
\label{den}\ee where $T_\ga = 2.4 \times 10^{-4}$ eV is the
present photon temperature, $g\left(T_\ga\right) = 2$ is the
photon degree of freedom and $\rho_c = 7.5 \times 10^{-47}h^2$,
with $h = 0.71$, being the critical density of the Universe. Let
us take $m_h = 7.75$ MeV, $v = 174$ GeV, $a_5 =0.65$, $-a_6 =
0.38$ (actually in our calculations, we have used a better
precision for $a_5$ and $a_6$) and $m_1 = 150$ GeV. Thus, from
Eqs. (\ref{beta}) and (\ref{den}) we obtain $\Om_h = 0.3$.
Therefore, without imposing any new fields or symmetries, the
3-3-1 model possesses a scalar field that can satisfy all the
properties required for the self-interacting dark matter and that
does not overpopulate the Universe. \par The candidate for
self-interacting dark matter that we propose here differs from the
singlet models of Refs. \cite{mcd,pos01} in an important point. As
we have discussed above it comes from a gauge model proposed with
another motivation that has an independent phenomenology.
Therefore, the values of the parameters that we impose here must
not spoil the preexisting bounds. We can obtain $m_1 \approx 150$
GeV from Eq. (\ref{h1}) with $a_1 = 1.2$ and $a_2 =0.36 $. From
Eq. (\ref{h3}) we have $m_3 = 1$ TeV. On the other hand, one
should notice that $m_h$ has a small value since $-f \sim
\left(10^{-7} - 10^{-6}\right)$ GeV and $u \sim 195$ GeV. However,
$h^0$ does not couple to the particles of the standard model for
the Higgs boson. Thus, it evades the present accelerator limits.
The constants $a_5$ and $a_6$ do not enter in the masses of the
particles of the model and so, it is free in this work
\cite{TAKL}. \par
\subsection{The 3-3-1 model with right-handed neutrinos}
 In this model the leptons are in triplets, and the third member
is a RH neutrino: \be f^{a}_L = \left(
               \nu^a_L, e^a_L, (\nu^c_L)^a
\right)^T \sim (1, 3, -1/3), e^a_R\sim (1, 1, -1). \label{l2} \ee
 Each charged left-handed fermion field has its right-handed counterpart
transforming as a singlet of the SU(3)$_L$ group. The first two
generations of quarks are in antitriplets while the third one is
in a triplet: \be Q_{iL} = \left(
                d_{iL},-u_{iL}, D_{iL}\\
                 \right)^T \sim (3, \bar{3}, 0),
\label{q} \ee
\[ u_{iR}\sim (3, 1, 2/3), d_{iR}\sim (3, 1, -1/3),
D_{iR}\sim (3, 1, -1/3),\ i=1,2,\] \be
 Q_{3L} = \left(
                 u_{3L}, d_{3L}, T_{L}
                 \right)^T \sim (3, 3, 1/3),
\ee
\[ u_{3R}\sim (3, 1, 2/3), d_{3R}\sim (3, 1, -1/3), T_{R}
\sim (3, 1, 2/3).\] The charged gauge bosons are defined as \bea
\sqrt{2}\ W^+_\mu &=& W^1_\mu - iW^2_\mu ,
\sqrt{2}\ Y^-_\mu = W^6_\mu - iW^7_\mu ,\nn\\
\sqrt{2}\ X_\mu^o &=& W^4_\mu - iW^5_\mu. \eea \hs  The {\it
physical} neutral gauge bosons are again related to $Z, Z'$
through the mixing angle $\phi$. The symmetry breaking can be
achieved with just three $\mbox{SU}(3)_L$ triplets
\bea \chi & =
&\left(
                \chi^0, \chi^-, \chi^{,0}
                 \right)^T \sim (1, 3, -1/3),\nn\\
\rho & =& \left(
                \rho^+, \rho^0, \rho^{,+}
                  \right)^T \sim (1, 3, 2/3),\label{hig}\\
\eta & =& \left(
                \eta^0, \eta^-, \eta^{,0}
                 \right)^T \sim (1, 3, -1/3),\nn.
\eea The necessary VEVs are \be \langle\chi \rangle = (0, 0,
\om/\sqrt{2})^T,\ \langle\rho \rangle = (0, u/\sqrt{2}, 0)^T,\
\langle\eta \rangle = (v/\sqrt{2}, 0, 0)^T. \label{vev} \ee
 After symmetry
breaking the gauge bosons gain  masses \be
m^2_W=\frac{1}{4}g^2(u^2+v^2),\ M^2_Y=\frac{1}{4}g^2(v^2+\om^2),
M^2_X=\frac{1}{4}g^2(u^2+\om^2). \label{rhb} \ee Eqn.(\ref{rhb})
gives us a relation \be v_W^2 = u^2 + v^2 =246^2\ \  \mbox{GeV}^2.
\label{vw} \ee \hs  In order to be consistent with the low energy
phenomenology we have to assume that $\langle \chi \rangle \gg\
\langle \rho \rangle,\ \langle \eta \rangle$ such that $m_W \ll
M_X, M_Y$. \hs  The symmetry-breaking hierarchy gives us splitting
on the bilepton masses~\cite{til} \be | M_X^2 - M_Y^2 | \leq
m_W^2. \label{mar} \ee \hs Our aim in this paper is to show that
the 3-3-1 model with RH neutrinos furnishes a good candidate for
(self-interacting) dark matter. The main properties that a good
dark matter candidate must satisfy are stability and neutrality.
Therefore, we go to the scalar sector of the model, more
specifically to the neutral scalars, and we examine whether any of
them can be stable and in addition whether they can satisfy the
self-interacting dark matter criterions \cite{ss}. In addition,
one should notice that such dark matter particle must not
overpopulate the Universe. On the other hand, since our dark
matter particle is not imposed arbitrarily to solve this specific
problem, we must check that the necessary values of the parameters
do not spoil the other bounds of the model. \hs Under assumption
of the discrete symmetry  $\chi \rightarrow - \chi$, the most
general potential can then be written in the following
form~\cite{l97} \bea V(\eta,\rho,\chi)&=&\mu^2_1 \eta^+ \eta +
 \mu^2_2 \rho^+ \rho +  \mu^2_3 \chi^+ \chi +
\lambda_1 (\eta^+ \eta)^2 + \lambda_2 (\rho^+ \rho)^2 + \lambda_3
(\chi^+ \chi)^2 \crn & + & (\eta^+ \eta) [ \lambda_4 (\rho^+ \rho)
+ \lambda_5 (\chi^+ \chi)] + \lambda_6 (\rho^+ \rho)(\chi^+ \chi)
+
\lambda_7 (\rho^+ \eta)(\eta^+ \rho) \nonumber\\
& + & \lambda_8 (\chi^+ \eta)(\eta^+ \chi) + \lambda_9 (\rho^+
\chi)(\chi^+ \rho) +
 \lambda_{10} (\chi^+ \eta + \eta^+ \chi)^2.
\label{pot1} \eea We rewrite the expansion of the scalar fields
which acquire a VEV: \be \eta^o = \fr{1}{ \sqrt{2}}\left(v +
\xi_\eta + i \zeta_\eta\right) ; \ \rho^o =  \fr{1}{
\sqrt{2}}\left( u + \xi_\rho + i \zeta_\rho\right);\ \chi^o =
\fr{1}{ \sqrt{2}}\left( w + \xi_\chi + i \zeta_\chi\right).
\label{exp1} \ee
 For the  prime neutral fields which do
not have VEV, we get analogously: \be \eta'^{o} = \fr{1}{
\sqrt{2}}\left( \xi'_\eta + i \zeta'_\eta\right) ; \ \chi'^{o} =
\fr{1}{ \sqrt{2}}\left( \xi'_\chi + i \zeta'_\chi\right).
\label{exp2} \ee Requiring that in the shifted potential $V$, the
linear terms in fields must be absent, we get  in the tree level
approximation,
 the following constraint equations:
\bea \mu^2_1 +  \la_1 v^2 +\frac 1 2 \la_4 u^2 +\frac 1 2 \la_5
w^2
 & = & 0, \nonumber\\
\mu^2_2 +  \la_2 u^2 + \frac 1 2 \la_4 v^2 +\frac 1 2 \la_6 w^2
 & = & 0,
\label{cont}\\
\mu^2_3 +  \la_3  w^2 +\frac 1 2  \la_5 v^2 + \frac 1 2 \la_6 u^2
 & = & 0\nn .
\eea
\hs Since dark matter has to be neutral, then we consider
 only neutral Higgs
sector. In the $\xi_\eta, \xi_\rho, \xi_\chi, \xi'_\eta,
\xi'_\chi$ basis the square mass matrix, after imposing of the
constraints ~(\ref{cont}), has a quasi-diagonal form as follows:
\be M^2_H = \left( \begin{array}{cc}
M^2_{3H}& 0\\
0 & M^2_{2H} \end{array} \right), \ee where \be M^2_{3H} = \fr 1 2
\left( \begin{array}{ccc}
 2 \la_1 v^2  & \la_4 v u   &
  \la_5 v w   \\
  \la_4 v u   &  2 \la_2 u^2 &
  \la_6 u w \\
  \la_5 v w  &  \la_6 u w  &
 2 \la_3 w^2  \end{array} \right),
\label{mat1} \ee and \be M^2_{2H} = \left( \fr{\la_8}{4} +
\la_{10} \right)\Biggl(
\begin{array}{cc}
  w^2    & v w \\
 v w & v^2 \end{array} \Biggr).
\label{mat2} \ee The above mass matrix shows that the prime fields
mix themselves but do not mix  with others. In the limit \be
\la_1v,\ \la_2 u,\ \la_4 u \ll \la_5 w,\ \la_6 w, \label{lm1} \ee
we obtained physical eigenstates  $H_1(x)$ and   $\si(x)$ \be
\Biggl(
\begin{array}{c}
  H_1(x)\\
 \si(x) \end{array} \Biggr)
= \frac{1}{(\la_5^2 v^2 + \la_6^2 u^2)^{1/2}} \Biggl(
\begin{array}{cc}
  \la_6 u   & -\la_5 v \\
 \la_5 v & \la_6 u \end{array} \Biggr)
\Biggl( \begin{array}{c}
  \xi_\eta\\
 \xi_\rho \end{array} \Biggr),
\label{ct17} \ee with masses~\cite{l97} \bea m^2_{H_1}& \approx&
\frac{v^2}{4\la_6}( 2 \la_1 \la_6 -  \la_4 \la_5) \approx
 \frac{u^2}{4\la_5}(
2 \la_2 \la_5 -  \la_4 \la_6), \label{rt1}
\\
m^2_{\si} &\approx&  \fr 1 2 \la_1 v^2 +  \frac{\la_4 \la_6
u^2}{4\la_5} \approx
 \fr 1 2 \la_2 u^2 +  \frac{\la_4
\la_5 v^2}{4\la_6}. \label{rt2} \eea Eqs.~(\ref{rt1})
and~(\ref{rt2}) also give us relations among coupling constants
and VEVs. Another  massive physical state $H_3$ with mass: \be
  m^2_{H_3} \approx -  \la_3 w^2 .
\label{mass1} \ee The scalar $\si(x)$ is the one that we can
identify with the SM Higgs boson~\cite{l97}. \hs In the
approximation $w\gg v$,
 mass matrix $M^2_{2H}$
gives us one Goldstone $\xi'_\chi$ and one physical massive field
$\xi'_\eta$ with mass \be m^2_{\xi'_\eta} =  -\left( \fr{\la_8}{4}
+  \la_{10}\right) w^2. \label{mass2} \ee
 In  the pseudoscalar  sector, we have three Goldstone  bosons
which can be identified as follows: $ G_2 \equiv \zeta_\eta,\ G_3
\equiv  \zeta_\rho,\ G_4 \equiv  \zeta_\chi$ and in the
$\zeta'^o_\eta, \zeta'^o_\chi$ basis \be M^2_{2A} = \left(
\fr{\la_8}{4} +  \la_{10} \right)\Biggl(
\begin{array}{cc}
  w^2    & v w \\
 v w & v^2 \end{array} \Biggr).
\label{mat3} \ee We easily  get one Goldstone  $G'_5$ and one
massive pseudoscalar boson $\zeta'_\eta$ with mass \be
m^2_{\zeta'_\eta} =  -\left( \fr{\la_8}{4} +  \la_{10}\right) w^2.
\ee It is to be emphasized that, both $\xi'_\eta$ and
$\zeta'_\eta$
 are in  an  singlet of the $SU(2)$. Therefore they do not interact
with the SM gauge bosons $W^\pm, Z^0\ \mbox{and}\ \ga$. Unlike the
3-3-1 model considered in~\cite{ft}, here we have two fields which
can be considered as dark matter.
 To get the interaction of dark matter to the SM Higgs boson,
we consider the following relevant parts \bea L_{int}(\si,
\zeta_\eta) &=& \fr 1 4 \la_1\left[ v^2 + 2v \xi_\eta + \xi_\eta^2
+ \zeta_\eta^2
+ \xi_\eta^{'2} + \zeta_\eta^{'2} +2\eta^+\eta^- \right]^2 \nn\\
&+& \fr 1 4 \la_4\left[ v^2 + 2v \xi_\eta + \xi_\eta^2 +
\zeta_\eta^2
+ \xi_\eta^{'2} + \zeta_\eta^{'2} +2\eta^+\eta^- \right]\nn\\
&\times&\left[ u^2 + 2u \xi_\rho + \xi_\rho^2 + \zeta_\rho^2 +
2\rho^+\rho^- + 2\rho^{'-}\rho^{'+}\right] \label{shtt} \eea
Substituting (\ref{ct17}) we get  couplings of SIDM with the SM
Higgs boson $\si$ \bea L(\si, \zeta_\eta)&=&\left[
\fr{\si(x)}{\sqrt{\la_5^2 v^2+\la_6^2 u^2}} \left(\la_1 \la_5 v^2
+ \fr{\la_4 \la_6}{2}u^2\right) + \fr{H_1(x)\si(x)}{ (\la_5^2
v^2+\la_6^2 u^2)}\left(\la_1 -\fr{\la_4}{2}\right)
\la_5 \la_6 uv \right.\nn\\
&+& \left. \fr{\si^2(x)}{2(\la_5^2 v^2+\la_6^2 u^2)} \left(
\la_5^2 v^2 +\fr{\la_6^2}{2} u^2\right)\right]\left(\xi_\eta^{'2}
+\zeta_\eta^{'2}\right). \label{sdm} \eea From Yukawa couplings,
we see that our candidates do not interact with ordinary leptons
and quarks~\cite{l96}. \bea {\cal
L}_{Yuk}^{\eta}&=&\la_{3a}\bar{Q}_{3L}u_{aR}\eta+
\la_{4ia}\bar{Q}_{iL}d_{aR}\eta^{*}+\mbox{h.c.}\nonumber\\
&=&\la_{3a}(\bar{u}_{3L}\eta^o+\bar{d}_{3L}\eta^-+\bar{T}_L\eta^{,o})
u_{aR}+\la_{4ia}(\bar{d}_{iL}\eta^{o*}-\bar{u}_{iL}\eta^+
+\bar{D}_{iL}\eta^{,o*})d_{aR}+\mbox{h.c.}\nonumber \eea We see
that the candidates for dark matter in this model have not
couplings with all the SM particles except for the Higgs boson.
\hs For stability of  DM, we have to put mass of the SM Higgs
boson is twice bigger  mass of the candidate \be
 \fr 1 2 \la_1 v^2 +  \frac{\la_4
\la_6 u^2}{4\la_5} \approx
 \fr 1 2 \la_2 u^2 +  \frac{\la_4
\la_5 v^2}{4\la_6} \ge  -\left( \fr{\la_8}{4} + \la_{10}\right)
w^2. \ee To avoid the interaction of DM with Goldstone boson, we
have \be \la_1 = \fr{\la_4}{2} \ee \hs  The {\it wrong} muon decay
($\mu^- \rightarrow e^- \nu_e {\bar \nu}_\mu$) gives a lower limit
for singly charged bilepton $ M_Y   \sim 230 \ \mbox{GeV}$.
 Combining Eqns. (\ref{rhb},
\ref{vw}) with (\ref{mar}) we obtain the following relation: $u \
\sim \ v \ \approx 100 - 200 \ \mbox{GeV} $  and  $w \approx (500
- 1000)$ GeV. The cross section for $h h \rightarrow h h $ (where
$h$ stands for $\xi'_\eta$ and $\zeta'_\eta$) with quartic
interaction is $\si = \la_1^2/4 \pi m_h^2$. The requirement on the
quality $\si_{el}/(m_h[GeV])$ denoting the ration of the
 DM  elastic cross section to its  mass (measured in GeV)
is that~\cite{ss,mcd,far00}
 \be   2.05 \times 10^3 \ \mbox{GeV}^{-3}
 \leq \fr{\si}{m_{h}} \leq  2.57 \times 10^4 \
 \mbox{GeV}^{-3} \label{req} \ee Taking
$\la_1=1$  we get  4.7  MeV $\le m_{h} \ \le 23 $ MeV. The SIDM
candidates interact with the SM Higgs boson by strength 0.65 if
$\la_5 = \la_6 =1$ and $u = v = 175$ GeV are taken.
  Now consider the cosmic density of the $h$ scalar given by
~\cite{ft}: \be \Om_h=2g(T_\ga)T_\ga^3 \fr{m_h\bet}{\rho_cg(T)},
\ee
 where $T_\ga=2.4\times 10^{-4}$ eV
is the present photon temperature, $g(T_\ga) = 2$ is the photon
degree of freedom and $\rho_c=7.5 \times10^{-47} h^2$ with $
h=0.71 $, being the critical density of the Universe.  Taking $
m_h = 4.7$ MeV,  we obtain $\Om_h=0.18$. This means that the SIDM
candidates do not overpopulate the Universe.
\subsection{The 3-3-1 model with exotic neutral lepton}
In this model the leptons are in triplets, and the third member is
a {\it neutral exotic} lepton and left-handed fermion field has
its right-handed counterpart transforming as a singlet of the
SU(3)$_L$ group ~\cite{chen}: \be f^{a}_L = \left(
               \nu^a_L, e^a_L, N_L^a
\right)^T \sim (1, 3, -1/3), e^a_R\sim (1, 1, -1), N^a_R\sim (1,
1, 0). \label{l2} \ee The first two generations of quarks are in
antitriplets while the third one is in a triplet: \be Q_{iL} =
\left(
                d_{iL},-u_{iL}, D_{iL}\\
                 \right)^T \sim (3, \bar{3}, 0),
\label{q} \ee
\[ u_{iR}\sim (3, 1, 2/3), d_{iR}\sim (3, 1, -1/3),
D_{iR}\sim (3, 1, -1/3),\ i=1,2,\] \be
 Q_{3L} = \left(
                 u_{3L}, d_{3L}, T_{L}
                 \right)^T \sim (3, 3, 1/3),
\ee
\[ u_{3R}\sim (3, 1, 2/3), d_{3R}\sim (3, 1, -1/3), T_{R}
\sim (3, 1, 2/3).\] The charged gauge bosons are defined as \bea
\sqrt{2}\ W^+_\mu &=& W^1_\mu - iW^2_\mu ,
\sqrt{2}\ Y^-_\mu = W^6_\mu - iW^7_\mu ,\nn\\
\sqrt{2}\ X_\mu^o &=& W^4_\mu - iW^5_\mu. \eea \hs  The {\it
physical} neutral gauge bosons are again related to $Z, Z'$
through the mixing angle $\phi$. The symmetry breaking can be
achieved with  three $\mbox{SU}(3)_L$ triplets and an sextet \bea
\chi & = &\left(
                \chi^0, \chi^-, \chi^{,0}
                 \right)^T \sim (1, 3, -1/3),\nn\\
\rho & =& \left(
                \rho^+, \rho^0, \rho^{,+}
                  \right)^T \sim (1, 3, 2/3),\label{hig}\\
\eta & =&\left(
\begin{array}{ccc}
\si _{_{1}}^{o} & \frac{h_{_{1}}^{-}}{\sqrt{2}} &
\frac{\si_2^0}{\sqrt{2}} \crn \frac{h_{_{1}}^{-}}{\sqrt{2}} &
H_{_{1}}^{--} & \frac{h_{_{2}}^{-}}{\sqrt{2}} \crn
\frac{\si_2^0}{\sqrt{2}} & \frac{h_{_{1}}^{-}}{\sqrt{2}} &
\si_3^0\nn
\end{array}
\right)\sim (1,6,- 2/ 3). \eea
 The necessary VEVs are \be \langle\chi \rangle = (0, 0,
\om/\sqrt{2})^T,\ \langle\rho \rangle = (0, u/\sqrt{2}, 0)^T,\
\langle\eta \rangle = (v/\sqrt{2}, 0, 0)^T. \label{vev} \ee The
VEV of sextet has the form
 \bea \langle S\rangle =\left(
\begin{array}{ccc}
0 & 0 & v_{\si_2}
 \crn
0 &0 &0
 \crn
v_{\si_2} &0 & 0\nn
\end{array}
\right) \eea
 After symmetry
breaking the gauge bosons gain  masses \be
m^2_W=\frac{1}{4}g^2(u^2+v^2 +v_4),\
M^2_Y=\frac{1}{4}g^2(v^2+\om^2 +v_4^2),
M^2_X=\frac{1}{4}g^2(u^2+\om^2 +v_4^2). \label{rhb} \ee
Eqn.(\ref{rhb}) gives us a relation \be v_W^2 = u^2 + v^2 +v_4^2
=246^2\ \  \mbox{GeV}^2. \label{vw} \ee \hs  In order to be
consistent with the low energy phenomenology we have to assume
that $\langle \chi \rangle \gg\ \langle \rho \rangle,\ \langle
\eta \rangle$ such that $m_W \ll M_X, M_Y$. \hs  The
symmetry-breaking hierarchy gives us splitting on the bilepton
masses~\cite{til} \be | M_X^2 - M_Y^2 | \leq m_W^2. \label{mar}
\ee The Yukawa couplings are given \be {\cal L}_Y = f_1 {\bar L}_a
\phi \nu_R + \fr 1 2 f_1 {\bar L}_a S L_b  \ee The mass
eigenstates ($\nu_e, \tilde{N}$) are related to the weak
eigenstates  ($\nu_e, \tilde{N}$) \be \tilde{\nu} = \nu \cos
\theta - N \sin \theta , \tilde{N} = \nu \sin \theta + N \cos
\theta \ee where $\theta$ is a mixing angle. A mass distribution
extended well beyond the visible galaxy ~\cite{gala} \be M(R) =
\int^R \rho (r) dV = v^2 R G^{-1} \ee
 \bc
\begin{picture}(350,110)(0,0)
\ArrowLine(10,30)(50,30)
  \ArrowLine(50,30)(90,30)
  \ArrowLine(90,30)(130,30)
  \PhotonArc(70,30)(20,0,180){2}{10}
  \Photon(70,50)(70,80){2}{3}
  \Text(10,20)[]{$\nu_e$}
    \Text(80,20)[]{$e$}
   \Text(120,20)[]{$\nu_e$}
 \Text(45,50)[]{$W^+$}
  \Text(100,50)[]{$W^-$}
  \Text(85,80)[]{$\ga$}
\ArrowLine(230,30)(270,30)
 \Photon(270,30)(310,30){2}{5}
 \ArrowLine(310,30)(350,30)
 \ArrowLine(270,30)(290,70)
 \ArrowLine(290,70)(310,30)
 \Photon(290,70)(290,100){2}{3}
 \Text(240,20)[]{$\nu_e$}
  \Text(290,20)[]{$W^+$}
   \Text(340,20)[]{$\nu_e$}
 \Text(270,60)[]{$e^-$}
  \Text(315,60)[]{$e^-$}
  \Text(300,90)[]{$\ga$}
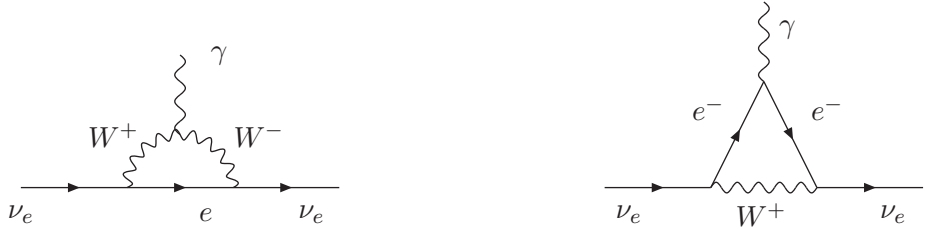
\Text(200,-7)[]{Fig.1: Two diagrams for the decay $N \rightarrow
\nu_e
 \ga$ }
\end{picture}
\ec \vs
\bc
\begin{picture}(350,110)(0,0)
\ArrowLine(10,30)(50,30)
  \ArrowLine(50,30)(90,30)
  \ArrowLine(90,30)(130,30)
  \DashCArc(70,30)(20,0,180){5}
  \Photon(70,50)(70,80){2}{3}
  \Text(10,20)[]{$\nu_e$}
    \Text(80,20)[]{$e$}
   \Text(120,20)[]{$\nu_e$}
 \Text(45,50)[]{$H$}
  \Text(100,50)[]{$H$}
  \Text(85,80)[]{$\ga$}
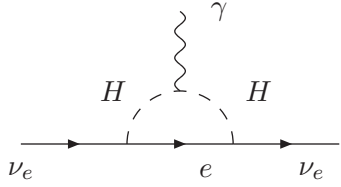
\Text(200,-7)[]{Fig.2:  Diagram for Higgs contribution to the
decay $N \rightarrow \nu_e
 \ga$ }
\end{picture}
\ec \vs
 The decay of $N$ particle into neutrino and photon contains
three graphs depicted in Fig.1 which give~\cite{chen} \be \Ga (N
\rightarrow \nu_e \ga) = \fr 9 4 \fr{m_N^5 G_F^2 \al}{512\ \pi^2}
\sin^2(2\theta) \label{rate}\ee which corresponds to lifetime \be
\tau \approx 4.67 \times 10^{14} \sin^{-2}(2\theta) ( 1
\mbox{keV}/m_N c^2)^5 \mbox{yr} \ee It is obvious that the $N$-
particle's life can be longer than the age of the universe, if
$m_N \sim 1$ kEV. It was shown that the Higgs contribution is much
smaller than those of the two first, and the $N$- particle meets
the constraints on dark matter from cosmology and galaxy
formation.
\section{Supersymmetric 3-3-1 model with right-handed neutrinos}
\hs Supersymmetric version on the 3-3-1 model with right-handed
neutrinos has been proposed in~\cite{331susy,sup}.  Here we will
follow the usual notation writing for a given fermion $f$, the
respective sfermions by $\tilde{f}$ {\it i.e.}, $\tilde{l}$ and
$\tilde{q}$ denote sleptons and squarks respectively. Then, we
have the following additional representations \bea \tilde{Q}_{\al
L}& =&
\left( \begin{array}{c} \tilde{d}_\al \\
                        \tilde{u}_\al \\
                        \tilde{d}^{ \prime}_\al
         \end{array}\right)_L
\sim ( {\bf 3},{\bf 3^{*}},0 ), \; \tilde{Q}_{3L} =
\left( \begin{array}{c} \tilde{u}_3 \\
                        \tilde{d}_3 \\
                        \tilde{u}^{ \prime}
         \end{array}
\right)_{L}\!\!\!\! \sim ( {\bf 3},{\bf 3},1/3 ), \crn
\tilde{L}_{aL} &=&
\left( \begin{array}{c} \tilde{ \nu}_{a} \\
                        \tilde{l_a} \\
                        \tilde{ \nu}^{c}_{a}
\end{array} \right)_{L} \sim ( {\bf 1},{\bf 3},-1/3 ),
\label{susytri} \eea \bea \lefteqn{\tilde{l}^{c}_{aL}\sim ({\bf
1},{\bf 1},1),} \crn && \tilde{u}^{c}_{i L},\,\tilde{u}^{ \prime
c}_{L} \sim ( {\bf 3}^*,{\bf 1},-2/3),\; \tilde{d}^{c}_{iL},
\tilde{d}^{ \prime c}_{\al L}
 \sim ( {\bf 3}^*,{\bf 1},1/3),
\label{ss} \eea with $a=e, \mu , \tau$; $i=1,2,3$; and $\al=1,2$.
However, when considering quark (or squark) singlets of a given
charge we will use the notation $u^c_{iL},d^c_{iL}$
($\tilde{u}_{iL},\tilde{d}^c_{iL}$ with $i(j)=1,2,3$. The
supersymmetric partner of the scalar Higgs fields, the higgsinos,
are \bea \tilde{ \eta} &=&
\left( \begin{array}{c} \tilde{ \eta}^{0}_{1} \\
                        \tilde{ \eta}^{-} \\
                        \tilde{ \eta}^{0}_{2}
\end{array} \right),\;
\tilde{ \chi} =
\left( \begin{array}{c} \tilde{ \chi}^{0}_{1} \\
                        \tilde{ \chi}^{-} \\
                        \tilde{ \chi}^{0}_{2}
\end{array} \right) \sim ( {\bf 1},{\bf 3},-1/3 ),
\crn \tilde{ \rho} &=&
\left( \begin{array}{c} \tilde{ \rho}^{+}_{1} \\
                        \tilde{ \rho}^{0} \\
                        \tilde{ \rho}^{+}_{2}
\end{array} \right) \sim ( {\bf 1},{\bf 3},2/3 ),
\label{esca} \eea and the respective extra higgsinos, needed to
cancel the chiral anomaly of the higgsinos in Eq.~(\ref{esca}),
are \bea \tilde{\eta}^{\prime} &=&
\left( \begin{array}{c} \tilde{\eta}^{\prime 0}_{1} \\
                        \tilde{\eta}^{\prime +} \\
                        \tilde{\eta}^{\prime 0}_{2}
         \end{array} \right),
\tilde{\chi}^{\prime} =
\left( \begin{array}{c} \tilde{ \chi}^{\prime 0}_{1} \\
                        \tilde{ \chi}^{\prime +} \\
                        \tilde{ \chi}^{\prime 0}_{2}
\end{array} \right) \sim ( {\bf 1},{\bf 3^{*}},1/3 ),
\crn \tilde{\rho}^{\prime} &=&
\left( \begin{array}{c} \tilde{\rho}^{\prime -}_{1} \\
                        \tilde{\rho}^{\prime 0} \\
                        \tilde{\rho}^{\prime -}_{2}
\end{array} \right) \sim ( {\bf 1},{\bf 3^{*}},-2/3 ),
\label{escac} \eea and the corresponding scalar partners denoted
by $\eta^{\prime}$,$\chi^{\prime}$, $\rho^{\prime}$, with the same
charge assignment as in Eq.~(\ref{escac}), and with the following
VEVs: $v^{\prime}=\langle \eta^{\prime 0}_1 \rangle/\sqrt2$,
$w^{\prime}=\langle \chi^{\prime 0}_2\rangle /\sqrt2$ and $
u^{\prime}=\langle \rho^{\prime 0} \rangle/\sqrt2$. This complete
the representation content of this supersymmetric model.
Concerning the gauge bosons and their superpartners, if we denote
the gluons by $g^b$ the respective superparticles, the gluinos,
are denoted by $\la^b_{C}$, with $b=1, \ldots,8$; and in the
electroweak sector we have $V^b$, the gauge boson of $SU(3)_{L}$,
and their gauginos partners $\la^b_{A}$; finally we have the gauge
boson of $U(1)_{N}$, denoted by $V^\prime$, and its supersymmetric
partner $\la_{B}$. This is the total number of fields in the
minimal supersymmetric extension of the 3-3-1 model of
Refs.~\cite{flt,pt93}.
\subsection{Superfields}
The superfields formalism is useful in writing the Lagrangian
which is manifestly invariant under the supersymmetric
transformations~\cite{wb} with fermions and scalars put in chiral
superfields while the gauge bosons in vector superfields. As usual
the superfield of a field $\phi$ will be denoted by
$\hat{\phi}$~\cite{mssm}. The chiral superfield of a multiplet
$\phi$ is denoted by \bea
\hat{\phi}\equiv\hat{\phi}(x,\theta,\bar{\theta})&=&
\tilde{\phi}(x) + i \; \theta \si^{m} \bar{ \theta} \;
\partial_{m} \tilde{\phi}(x) +\frac{1}{4} \; \theta \theta \;
\bar{ \theta}\bar{ \theta} \; \Box \tilde{\phi}(x) \crn & &
\mbox{} +  \sqrt{2} \; \theta \phi(x) + \frac{i}{ \sqrt{2}} \;
\theta \theta \; \bar{ \theta} \bar{ \si}^{m}
\partial_{m}\phi(x)
\crn && \mbox{}+  \theta \theta \; F_{\phi}(x), \label{phi} \eea
while the vector superfield is given by \bea
\hat{V}(x,\theta,\bar\theta)&=&-\theta\si^m\bar\theta V_m(x)
+i\theta\theta\bar\theta
\overline{\tilde{V}}(x)-i\bar\theta\bar\theta\theta
\tilde{V}(x)\crn &+&\frac{1}{2}\theta\theta\bar\theta\bar\theta
D(x). \label{vector} \eea The fields $F$ and $D$ are auxiliary
fields which are needed to close the supersymmetric algebra and
eventually will be eliminated using their motion equations. For
fermion superfields we use the notation \be \hat{L}_{aL},\;
\hat{l}^{c}_{aL},\; \hat{Q}_{\al L},\; \hat{Q}_{3L},\;
\hat{u}^{c}_{iL},\; \hat{d}^{c}_{iL},\; \hat{u}^{\prime c}_{L},\;
\hat{d}^{\prime c}_{\al L}. \label{superfermions} \ee For scalar
superfields we write: $\hat{ \eta}, \; \hat{ \chi},\;\hat{ \rho}$
and similar expressions for $\hat{\eta}^{\prime}$,
$\hat{\chi}^{\prime}$, $\hat{\rho}^{\prime}$ and we must change
$\mbox{(field)}$ by $\mbox{(field)}^{\prime}$. The vector
superfield for the gauge bosons of each factor $SU(3)_C$,
$SU(3)_L$ and $U(1)_N$ are denoted by $\hat{V}_C,\hat{\bar V}_C$;
$\hat{V},\hat{\bar V}$; and $\hat{V^\prime}$, respectively, where
we have defined $\hat{V}_C=T^b\hat{V}^b_C$,
$\hat{V}=T^b\hat{V}^b$; $\hat{\bar V}_C=\bar{T}^b\hat{V^b}_C$,
$\hat{\bar V}=\bar{T}^b\hat{V^b}$; $T^b=\la^b/2$,
$\bar{T}^b=-\la^{*b}/2$ are the generators of triplet and
antitriplets representations, respectively, and $\la^b$ are the
Gell-Mann matrices. The Lagrangian of the model has the following
form \bea
   {\cal L}_{331S} &=& {\cal L}_{SUSY} + {\cal L}_{\mbox{soft}},
\label{lagra} \eea where ${\cal L}_{SUSY}$ is the supersymmetric
part and ${\cal L}_{\mbox{soft}}$ the soft terms breaking
explicitly the supersymmetry.
\subsection{The supersymmetric Lagrangian}
The supersymmetric part of the Lagrangian is decomposed  in the
lepton, quark, gauge, and the scalar sectors as follow: \bea {\cal
L}_{SUSY} &=&   {\cal L}_{\mbox{Lepton}}+ {\cal L}_{\mbox{Quark}}+
{\cal L}_{\mbox{Gauge}}+ {\cal L}_{\mbox{Scalar}},
\label{susyterm} \eea where \bea {\cal L}_{\mbox{Leptons}} &=&
\int\!\! d^{4}\theta\left[\,\hat{
\bar{L}}_{aL}e^{2g\hat{V}-\frac{g'}{3}\hat{V}'}
\hat{L}_{aL}+\hat{\bar{l}}^c_{aL}e^{g^\prime \hat{V}^\prime}
\hat{l}^c_{aL}\right], \label{lagrangiana1} \eea in the lepton
sector, we have omitted the sum over the three lepton family for
simplicity, and \bea {\cal L}_{\mbox{Quarks}} &=& \int\!\!
d^{4}\theta\!\!\left[ \hat{\bar{Q}}_{\al L}
e^{[2(g_s\hat{V}_{C}+g\hat{\bar{V}})]} \hat{Q}_{\al
L}+\hat{\bar{Q}}_{3L}
e^{[2(g_s\hat{V}_{C}+g\hat{V})+\frac{g'}{3}\hat{V}']} \hat{Q}_{3L}
 \,\right.
\crn&+& \left.\,\hat{ \bar{u}}^{c}_{iL} e^{[2g_s
\hat{\bar{V}}_C-\frac{2g'}{3}\hat{V}']} \hat{u}^{c}_{iL} +\hat{
\bar{d}^c}_{iL} e^{[2g_s \hat{ \bar{V}}_{C}+\frac{g'}{3}\hat{V}']}
\hat{d}^{c}_{iL}\right. \crn&+& \left.\,\hat{\bar{u}}^{\prime c}_L
e^{[2g_s \hat{ \bar{V}}_{C}-\frac{2g'}{3}\hat{V}']}
\hat{u}^{\prime c}_L +\hat{ \bar{d}^c}^{\prime }_{\al L} e^{[2g_s
\hat{ \bar{V}}_{C}+\frac{g'}{3}\hat{V}']} \hat{d}^{\prime c}_{\al
L} \right], \label{lagrangiana2} \eea in the quark sector, and we
have denoted $g_s,g,g^\prime$ the gauge coupling constants for the
$SU(3)_C,SU(3)_L,U(1)_N$ factors, respectively. In the gauge
sector we have \bea {\cal L}_{\mbox{Gauge}} &=&  \frac{1}{4} \int
d^{2}\theta\; \left[ \cal {W}_{C}\cal{W}_{C}+{\cal W}\cal{W}+{\cal
W}^{ \prime}{\cal W}^{ \prime}\,  \right] \crn &+& \frac{1}{4}
\int  d^{2}\bar{\theta}\; \left[ \bar{\cal W}_{C}\bar{\cal
W}_{C}+\bar{\cal W}\bar{\cal W}+ \bar{\cal W}^{ \prime}\bar{\cal
W}^{ \prime}\, \right], \label{lagrangiana3} \eea where
$\cal{W}_{C}$, $\cal{W}$ e $\cal{W}^{ \prime}$ are fields that can
be written as follows \bea \cal{W}_{\zeta C}&=&- \frac{1}{8g_s}
\bar{D} \bar{D} e^{-2g_s \hat{V}_{C}} D_{\zeta} e^{2g_s
\hat{V}_{C}},\crn \cal{W}_{\zeta}&=&- \frac{1}{8g} \bar{D} \bar{D}
e^{-2g \hat{V}} D_{\zeta} e^{2g \hat{V}}, \crn
\cal{W}^{\prime}_{\zeta}&=&- \frac{1}{4} \bar{D} \bar{D} D_{\zeta}
\hat{V}^{\prime}, \,\ \zeta=1,2. \label{cforca} \eea Finally, in
the scalar sector we have \bea {\cal L}_{\mbox{Escalar}} &=& \int
d^{4}\theta\;\left[\, \hat{ \bar{
\eta}}\,e^{[2g\hat{V}-\frac{g'}{3}\hat{V}']} \hat{ \eta} + \hat{
\bar{ \chi}}\,e^{[2g\hat{V}-\frac{g'}{3}\hat{V}']} \hat{ \chi} +
\hat{ \bar{ \rho}}\,e^{[2g\hat{V}+\frac{2g'}{3}\hat{V}']} \hat{
\rho} \right. \crn &+& \left.\,\hat{ \bar{
\eta}}^{\prime}\,e^{[2g\hat{ \bar{V}}+ \frac{g'}{3}\hat{V}']}
\hat{ \eta}^{\prime} + \hat{ \bar{ \chi}}^{\prime}\,e^{[2g\hat{
\bar{V}}+ \frac{g'}{3}\hat{V}']} \hat{ \chi}^{\prime} + \hat{
\bar{ \rho}}^{\prime}\,e^{[2g\hat{ \bar{V}}-
\frac{2g'}{3}\hat{V}']} \hat{ \rho}^{\prime} \right] \crn &+& \int
d^{2}\theta W+ \int d^{2}\bar{ \theta}\bar{W} \!,\hspace{2mm}
\label{escalagra} \eea where $W$ is the superpotential.
\subsection{The scalar Potential}
In the present model the scalar potential is written as \be
V_{331}=V_F+V_D+V_{\rm soft}, \label{p1} \ee where \bea
V_F&=&-{\cal L}_F=\sum_m F^\dagger_m F_m \crn
&=&\sum_{ijk}[\left\vert\frac{\mu_\eta}{2}\eta^\prime_i+
\frac{f_2}{3}\epsilon_{ijk}\rho_j\chi_k \right\vert^2+ \left\vert
\frac{\mu_\chi}{2}\chi^\prime_i+\frac{f_2}{3}\epsilon_{ijk}\eta_j\rho_k
\right\vert^2+ \left\vert
\frac{\mu_\rho}{2}\rho^\prime_i+\frac{f_2}{3}\epsilon_{ijk}\chi_j\eta_k
\right\vert^2\crn &+& \left\vert
\frac{\mu_\eta}{2}\eta_i+\frac{f^\prime_2}{3}\epsilon_{ijk}\rho^\prime_j
\chi^\prime_k\right\vert^2+ \left\vert
\frac{\mu_\chi}{2}\chi_i+\frac{f^\prime_2}{3}\epsilon_{ijk}
\eta^\prime_j\rho^\prime_k\right\vert^2+ \left\vert
\frac{\mu_\rho}{2}\rho_i+\frac{f^\prime_2}{3}\epsilon_{ijk}
\chi^\prime_j\eta^\prime_k\right\vert^2] \label{p2} \eea and \bea
V_D&=&-{\cal
L}_D=\frac{1}{2}(D^aD^a+DD)=\frac{g^{\prime2}}{18}(-\eta^\dagger\eta+
\eta^{\prime\dagger}\eta^\prime-\chi^\dagger\chi+\chi^{\prime\dagger}\chi^\prime
+2\rho^\dagger\rho-2\rho^{\prime\dagger}\rho^\prime)^2\crn &+&
\frac{g^2}{8}(\eta^\dagger_i\la^b_{ij}\eta_j-
\eta^{\prime\dagger}_i\la^{*b}_{ij}\eta^\prime_j+
\chi^\dagger_i\la^b_{ij}\chi_j-
\chi^{\prime\dagger}_i\la^{*b}_{ij}\chi^\prime_j+
\rho^\dagger_i\la^b_{ij}\rho_j+
\rho^{\prime\dagger}_i\la^{*b}_{ij}\rho^\prime_j)^2\!, \label{p3}
\eea finally, \bea V_{\rm soft}&=&-{\cal L}_{\rm soft}=
m^2_\eta\eta^\dagger\eta+m^2_\rho\rho^\dagger\rho+m^2_\chi\chi^\dagger\chi
+m^2_{\eta^\prime}\eta^{\prime\dagger}\eta^\prime \crn &+&
m^2_{\rho^\prime}\rho^{\prime\dagger}\rho^\prime+
m^2_{\chi^\prime}\chi^{\prime\dagger}\chi^\prime-
\epsilon_{ijk}(k_1 \rho_i\chi_j\eta_k+
k^\prime_1\rho^\prime_i\chi^\prime_j\eta^\prime_k\crn&+&H.c.),
\label{p4} \eea where we have used the scalar multiplets in
Eqs.~(\ref{vev}) and (\ref{escac}). With
Eqs.~(\ref{p2})-(\ref{p4}) we can work out the mass spectra of the
scalar and pseudoscalar fields by making the usual shift
$X^0\to\frac{1}{\sqrt2}(v_X+H_X+iF_X)$. The analysis is similar to
that of Ref.~\cite{331susy} and we will not write the constraints
equation, etc. By using as input the following values for the
parameters: $\sin^2\theta_W=0.2314$, $g=0.6532$,
$g^\prime=1.1466$; $f_2=2$, $f^\prime_2=10^{-3}$;
$k_1=k^\prime_1=10$ GeV; $\mu_\eta=\mu_\rho=\mu_\chi=-10^3$ GeV;
$m_\eta=15$ GeV, $m_\rho=10$ GeV. $m_\rho=244.99$ GeV;
$m_{\chi_2}=m_{\chi^\prime_2}=10^3$ GeV and $m_{\rho^\prime}=13$
GeV, we obtain the  masses \bea &&m_1\approx1702, m_2\approx1449,
m_3\approx387, \crn &&m_4\approx380,m_5\approx361, m_6\approx130,
\label{mi} \eea for the scalar sector (all masses are in GeV).
Note that the lightest neutral scalar is heavier than the lower
limit of the Higgs scalar of the standard model, i.e.,
$m_H\stackrel{>}{\sim} 114$ GeV. For the pseudoscalar sector we
obtain \bea &&M_1\approx1702, M_2\approx1449,M_3\approx363,\crn
 &&M_4\approx5, M_5=0, M_6=0,
\label{Mi} \eea only the two massless pseudoscalars are exact
values, i.e., there are two Goldstone bosons as it should be.
Notice that there is a light pseudoscalar. A carefully study shows
that there is a Higgs boson satisfies the conditions for
SIDM~\cite{llh}.
\section{Conclusions}
  In conclusion, it is a remarkable fact that the
3-3-1 model has an option for self-interacting dark matter without
the need of imposing any new symmetry to stabilize it. We have
shown that the Spergel-Steinhardt bound for self-interacting dark
matter \cite{pt93} can be realized in the 3-3-1 model with a
reasonable choice of the values of the parameters. The 3-3-1 model
with RH neutrinos provides two Higgs bosons: one is scalar or
$CP$-even and another is pseudoscalar or $CP$- odd particle having
properties of candidates for dark matter. In difference with the
previous candidate which introduced by hand, our self-interacting
dark matter arises without impose new properties to satisfy all
the criteria. From the conditions for SIDM one get the bounds for
scalar Higgs bosons: $m_h= 7.75 $ MeV in the 3-3-1 model with
exotic leptons~\cite{ft} and $4.7 \ {\mbox MeV} \leq m_h \leq 23 $
MeV in the 3-3-1 version with RH neutrinos~\cite{longlan}. Scalar
dark matter candidates have been recently investigated
in~\cite{bf}. It means that the candidate for SIDM has not to be
introduced {\it ad hoc} as in other models~\cite{sidm}. This
feature remains valid in the supersymmetric 3-3-1 model with
right-handed neutrinos. The spin $\fr 1 2$ exists in the version
with exotic neutral lepton. This would turn out to be particularly
interesting
 if observations reveal to be in favor of light
dark matter.
    This work was
supported in part by National Council for
Natural Sciences of Vietnam contract No: KT - 04.1.2.\\[0.3cm]


\begin{thebibliography}{9}
\bibitem{mt00} M. S. Turner, Phys. Rep. 333-334 (2000) 619.
\bibitem{bah99} N. A. Bahcall  {\it et al,} Science 284 (1999) 1481.
\bibitem{wan00} L. Wang  {\it et al,} Astrophys. J. 530 (2000) 17.
\bibitem{pos01} C. P. Burgess {\it et al,}
Nucl. Phys. B 619 (2001) 709; M. C. Bento  {\it et al,}
 Phys. Rev. D 62 (2000) 041302;
M. C. Bento  {\it et al,} Phys. Lett. B 518 (2000) 276; D. E. Holz
and A. Zee, Phys. Lett. B 517 (2001) 239; V. Silveira and A. Zee,
Phys. Lett. B 161 (1985) 136.
\bibitem{ghi} S. Ghigna  {\it et al,} Astrophys. J. 544 (2000) 616;
J. F. Navarro  {\it et al,} {\it ibid} 462 (1996) 563; B. Moore
{\it et al,} Mon. Not. R. Astron. Soc. 310 (1999) 1147.
\bibitem{moore} B. Moore  {\it et al,} Astrophys. J. 524 (1999) L19;
A. Klypin  {\it et al,} Astrophys. J. 522 (1999) 82.
\bibitem{dav01} R. Dav\'e  {\it et al,}
Astrophys. J., 547 (2001) 574.
\bibitem{bul00} J. S. Bullock  {\it et al,}
Astrophys. J. 539 (2000) 517; R. A. Swaters  {\it et al,}
Astrophys. J. 531 (2000) L107; F. C. van den Bosch  {\it et al,}
 Astrophys. J. 119 (2000) 1579;
J. F. Navarro and M. Steinmetz, Astrophys. J. 528 (2000) 607; C.
Firmani  {\it et al,} Mon. Not. R. Astron. Soc. 315 (2000) L29; V.
P. de Battista and J. A. Sellwood, Astrophys. J. 493 (1998) L5.
\bibitem{ss} D. N. Spergel and P. J. Steinhardt,
Phys. Rev. Lett. 84 (2000) 3760.
\bibitem{yoh00} N. Yoshida {\it et al.}, Astrophys. J. 544 (2000) L87.
\bibitem{buote} D. A. Buote  {\it et al,}
Astrophys. J. 577 (20020 183.
\bibitem{ost} J. P. Ostriker, Phys. Rev. Lett. 84 (2000) 5258;
 A. Burkert, astro-ph/0002409.
\bibitem{mcd} J. McDonald, Phys. Rev. Lett. 88 (2002) 091304.
\bibitem{ppf} F. Pisano and V. Pleitez, Phys. Rev.  D 46, (1992) 410;
P. H. Frampton, Phys. Rev. Lett.  69, (1992) 2889; R. Foot  {\it
et al,}
 Phys. Rev. D 47, (1993) 4158.
\bibitem{flt} R. Foot, H. N. Long, and Tuan A. Tran,
 Phys. Rev. D 50, (1994) R34;
H. N. Long, Phys. Rev. D 54, (1996) 4691.
\bibitem{ft} D. Fregolente and M. D. Tonasse, Phys. Lett. B555 (2003) 7.
\bibitem{longlan} H. N. Long and N. Q. Lan, Europhys. Lett. {\bf 64}, 571
(2003).
\bibitem{phf} P. H. Frampton and T. W. Kephart, Phys. Lett. {\bf B585}, 24
(2004); P. H. Frampton, Plenary Talk at SUGRA 20, Northeastern
University, March 2003, hep-th/0306029.
\bibitem{pl331} A. G. Dias, R. Martinez and V. Pleitez, hep-ph/0407074.
\bibitem{pr} C. A. de S. Pires and O. Ravinez, Phys. Rev. D {\bf58}, 035008
(1998).
\bibitem{cp} C. A. de S. Pires, Phys. Rev. D {\bf60}, 075013 (1999).
\bibitem{331susy}  H. N. Long and P. B. Pal, Mod. Phys. Lett. {\bf A13}, 2355
(1998);  T. V. Duong and E. Ma, Phys. Lett. {\bf B316}, 307
(1993); J. C. Montero, V. Pleitez and M. C. Rodriguez, Phys. Rev.
D {\bf 65}, 035006 (2002); R. A. Diaz, R. Ramirez, J. Mira and J.
A. Rodriguez, Phys. Lett {\bf B552}, 287 (2003).
\bibitem{marcos} M. Capdequi-Peyranere and M. C. Rodriguez, Phys. Rev. D {\bf
65}, 035006 (2002).
\bibitem{pq} R. D. Peccei and H. Quinn, Phys. Rev. Lett. {\bf38}, 1440 (1977);
Phys. Rev. D {\bf16}, 1791 (1977).
\bibitem{pal} P. B. Pal, Phys, Rev. D {\bf52}, 1659 (1995);
A. G. Dias, V. Pleitez, and M. D. Tonasse, Phys. Rev. D {\bf67},
095008 (2003); A. G. Dias and V. Pleitez, Phys. Rev. D{\bf 69},
015007 (2004); A. G. Dias, C. A. de S. Pires, and P. S. Rodrigues
da Silva ,Phys. Rev. D {\bf68}, 115009 (2003).
\bibitem{longvan} H. N. Long and V. T. Van, J. Phys. {\bf G25}, 2319 (1999).
\bibitem{f} R. Foot et al in~\cite{flt}.
\bibitem{pp95} F. Pisano and V. Pleitez, Phys. Rev. D {\bf51}, 3865
(1995).
\bibitem{rc} G. Tavares- Velasco and J. J. Toscano, Phys. Rev.
D 65, 013005 (2004);
\bibitem{pt93} V. Pleitez and M. D. Tonasse, Phys. Rev. D48,
(1993) 2353.
\bibitem{TAKL} M. D. Tonasse, Phys. Lett. B 381 (1996) 191
 [See also N. T. Anh, N. A. Ky and H. N. Long, Int. J. Mod. Phys.
 A 16 (2001) 541 and references cited therein].
\bibitem{pdg} Particle Data Group, S. Eidelman et al,
Phys. Lett. B592, 1 (2004).
\bibitem{kt}See for example, E. W. Kolb and M. S. Turner,
{\it The early universe} (Addison-Wesley Publishing Co., Reading,
1990).
\bibitem{til} H. N. Long and T. Inami,
 Phys. Rev. D {\bf61},
075002 (2000).
\bibitem{l97} H. N. Long, Mod. Phys. Lett. A 13 (1998) 1865.
\bibitem{l96} H. N. Long: Right-handed neutrino currents in the
$\mbox{SU(3)}_L\otimes\mbox{U(1)}_N$ electroweak theory.
 [hep-ph/9603258].
\bibitem{far00} B. D. Wandelt  {\it et al,} {\it in} Marina de Rey 2000,
Sources and detection of dark matter and dark energy in the
Universe, pp. 263 - 274, [astro-ph/0006344].
\bibitem{chen} F. Z. Chen, Phys. Lett. B 442 (1998) 223.
\bibitem{gala} A. De Rujula and S. L. Glashow, Phys. Rev.
Lett. 45 (1980) 942.
\bibitem{sup} J. C. Montero, V. Pleitez and M. C. Rodriguez,
{\it  Supersymmetric 3-3-1 model with right-handed neutrinos},
preprint IFT-P.027/2004 [ hep-ph/0406299].
\bibitem{wb} J. Wess e J. Bagger, {\it Supersymmetry and Supergravity}
2nd edition, Princeton University Press, Princeton NJ, (1992).
\bibitem{mssm} H. E. Haber and G. L. Kane, Phys. Rep. {\bf117}, 75 (1985).
\bibitem{llh} H. N. Long, N. Q. Lan and D. T. Huong,
work in progress.
\bibitem{bf} C. Boehm and P. Fayet, Nucl. Phys. B683 (2004) 219.
\bibitem{sidm} V. Silveira and A. Zee, Phys. Lett. {\bf B161},
136 (1985); D. E. Holz and A. Zee, Phys. Lett. {\bf B517}, 239
(2000); C. P. Burgess, M. Pospelov, and T. ter Veldhuis, Nucl.
Phys. {\bf B619}, 709 (2002); B. C. Bento, O. Bertolami, R.
Rosenfeld, and L. Teodoro, Phys. Rev. {\bf62}, 041302 (2000).
\end{thebibliography}
\end{document}